\documentclass[11pt,a4paper,fleqn]{article}
\usepackage[utf8]{inputenc}
\usepackage{authblk}
\usepackage{geometry}
\usepackage{xcolor}
\usepackage{natbib}
\usepackage{graphicx}
\usepackage{pdflscape}
\usepackage{appendix}

\usepackage{threeparttable}
\usepackage{booktabs}

\usepackage{amsmath}
\usepackage{bm}

\usepackage{enumitem}
\newlist{steps}{enumerate}{1}
\setlist[steps, 1]{label = Step \arabic*:}

\usepackage{dcolumn}
\newcolumntype{d}[1]{D{.}{.}{#1}}

\usepackage{caption}
\usepackage{subcaption}
\definecolor{nblue}{HTML}{000660}
\usepackage[colorlinks=true,urlcolor=nblue,linkcolor=nblue,citecolor=nblue]{hyperref}

\title{\textbf{Bayesian Inference in High-Dimensional Time-varying Parameter Models using Integrated Rotated Gaussian Approximations}}
\author[1]{\MakeUppercase{Florian Huber}\thanks{Corresponding author: Florian Huber. Salzburg Centre of European Union Studies, University of Salzburg. \textit{Address}: M\"{o}nchsberg 2a, 5020 Salzburg, Austria. \textit{Email}: \href{mailto:florian.huber@sbg.ac.at}{florian.huber@sbg.ac.at}. Florian Huber and Michael Pfarrhofer gratefully acknowledge financial support from the Austrian Science Fund (FWF, grant no. ZK 35).}}
\author[2]{\MakeUppercase{Gary Koop}}
\author[1]{\MakeUppercase{Michael Pfarrhofer}}   
\affil[1]{\textit{University of Salzburg}}
\affil[2]{\textit{University of Strathclyde}}
\date{\today}

\begin{document}

\maketitle
\begin{center}
\begin{minipage}{0.85\textwidth}
\begin{abstract}
\noindent Researchers increasingly wish to estimate time-varying parameter (TVP) regressions which involve a large number of explanatory variables. 
Including prior information to mitigate over-parameterization concerns has led to many using Bayesian methods. However, Bayesian Markov Chain Monte Carlo (MCMC) methods can be very computationally demanding. In this paper, we develop computationally efficient Bayesian methods for estimating TVP models using an integrated rotated Gaussian approximation (IRGA). This exploits the fact that whereas constant coefficients on regressors are often important, most of the TVPs are often unimportant. Since Gaussian distributions are invariant to rotations we can split the the posterior into two parts: one involving the constant coefficients, the other involving the TVPs. Approximate methods are used on the latter and, conditional on these, the former are estimated with precision using MCMC methods. In empirical exercises involving artificial data and a large macroeconomic data set, we show the accuracy and computational benefits of IRGA methods. 
\end{abstract}
\end{minipage}
\end{center}
\begin{center}
\begin{minipage}{0.7\textwidth}
\bigskip
\begin{tabular}{p{0.2\hsize}p{0.65\hsize}} 
\textbf{Keywords:}  & Time-varying parameter regression, Bayesian, Gaussian approximation, macroeconomic forecasting\\
\textbf{JEL Codes:} & C11, C30, E3, D31 \\
\end{tabular}
\end{minipage}
\end{center}
\normalsize

\newpage
\section{Introduction}
There has been an explosion of interest in carrying out structural analysis and forecasting with time-varying parameter (TVP) regressions and Vector Autoregressions (TVP-VARs) in recent years, with many of the papers employing Bayesian methods. An incomplete survey of significant recent Bayesian contributions includes \cite{cogley2005drifts},   \cite{primiceri2005}, \cite{ckls2012},  \cite{dh2012}, \cite{kk2012},  \cite{koop2013large},  \cite{giannone2013}, \cite{gpr2013}, \cite{nw2013},  \cite{bkk},  \cite{kg2014}, \cite{feldkircher2017sophisticated}, \cite{kmr2017},  \cite{ul2017}, \cite{koopkorobilisvb}, \cite{rm2018}, \cite{bitto_fs}, \cite{hauzenberger2019fast}, \cite{korobilis2019high}, \cite{paul2019timevarying} and \cite{hko2019}.

There are two issues in this literature that we wish to highlight as they form the basis of the contribution of our paper. The first is that Bayesian computation using Markov Chain Monte Carlo (MCMC) methods becomes increasingly computationally burdensome as the number of explanatory variables in the regression increases. Indeed, the contribution of much of the Big Data side of this literature lies in the development of computationally-efficient approximate methods which do not involve MCMC methods. 

The second issue is that the need for TVP methods seems to decrease as the number of explanatory variables increases. Many of the early contributions used a small number of regressors or involved low dimensional TVP-VARs and found strong evidence for the benefits of time-variation in coefficients. Indeed, the unobserved components stochastic volatility (UC-SV) model of \cite{sw2007} has had great success and it is the simplest of TVP regressions as it involves only an intercept. However, with more regressors, the benefits of allowing for time-variation in parameters has typically been found to be small.\footnote{This statement holds true for time-variation in regression coefficients. Allowing for stochastic volatility or other types of time-variation in the error variance is typically more important and leads to forecasting gains even in large models. In our macroeconomic forecasting exercise, every model allows for stochastic volatility.} A potential explanation for this finding is that, in TVP models with a small number of variables, the TVP aspect is controlling for omitted variables. When including a large number of explanatory variables, this factor is mitigated as the extra variables provide more heterogeneity that can be fit by the regression leaving less of a role for the TVPs.

 Accordingly, there may be many data sets where the researcher wishes to work with a TVP regression with many explanatory variables, but the computational burden using MCMC methods is prohibitive. This computational burden largely arises due to the TVP aspect of the model, and the researcher suspects that it is likely that most of the TVP part of the model is unnecessary. One alternative, of course, is to work with a constant coefficient regression model. But this risks missing small amounts of parameter change, perhaps only in a few parameters or in a few time periods, which are genuinely important. The methods developed in the present paper are designed for such a case. They are inspired by papers such as \cite{boom2019approximating} who use an integrated rotated Gaussian approximation (IRGA) which breaks the parameters in the model into two blocks. The first is for a low-dimensional set of regression coefficients which are of primary interest. The second is for a high-dimensional set of nuisance parameters. Fast and approximate methods are used with the latter and, conditional on the approximation obtained for the nuisance parameters, more accurate methods are used for the coefficients of interest. In our case, the nuisance parameters are the TVPs. 

To be precise, we exploit the rotation-invariance property of the Gaussian distribution to split the model into two parts: one involving constant coefficients and one involving the TVPs.  The first involves the explanatory variables with constant coefficients and a low-dimensional function of the TVPs. For this block we use MCMC methods (conditional on the approximation we use for the TVPs). The second block depends on the high-dimensional TVPs and for this we use a computationally fast approximation. In particular, we use computationally-fast vector approximate message passing (VAMP) methods, see \cite{rangan2019vector}.

Papers such as \cite{boom2019approximating}, focus on parameter estimation of the coefficients on the primary variables of interest as opposed to forecasting. We show that these methods can be extended to allow for efficient forecasting within a large TVP-VAR. Previous work with IRGAs assumed homoskedastic errors. An additional contribution of our paper lies in the development of methods for adding stochastic volatility (SV) to TVP regressions in the context of the IRGA approximation. 

Using artificial data, we demonstrate the computational scaleability and accuracy of IRGA methods. Moreover, employing a large US macroeconomic data set, we find IRGA estimation of TVP-VARs leads to good forecast performance in a computationally efficient manner. Finally, we use our large macroeconomic data set to investigate the effect of uncertainty shocks on the economy in a range of TVP-VARs of varying dimension. We find that evidence of time-variation in parameters decreases as the number of variables in the TVP-VAR increases. Our methods can discover this in a data based fashion in a manner in which either a constant coefficient VAR or a conventional TVP-VAR which allows for random walk behavior in all coefficients could not.  

The remainder of the paper is structured as follows: Section \ref{sec: rotation} discusses the static form of a TVP regression, provides details on the prior setup and shows how IRGA can be used to approximate the TVPs.  In Section \ref{sec: artData}, we use simulated data to illustrate the merits of the IRGA approach while Section \ref{sec: empApp} shows that our model also works well for a prominent US macroeconomic dataset. Finally, the last section summarizes and concludes the paper.

\section{Rotations in TVP Regression Models}\label{sec: rotation}

\subsection{The TVP Regression Model}

We write the TVP regression model as:\footnote{In the following discussion and without loss of generality, we normalize $\bm y$ and the regressors to have unit variance.}
\begin{align}
\bm y &= \bm X \bm \beta + \bm Z \bm \gamma + \bm \varepsilon, \label{eq: regmodel}
\end{align}
where $\bm y$ denotes a $T-$dimensional vector containing the dependent variable with typical element $y_t$ for $t=1,\dots,T$ and
$\bm X$ is a $T \times K$-dimensional vector of explanatory variables with typical $t^{th}$ row given by $\bm x_t'$ while $\bm \beta =(\beta_1, \dots, \beta_K)'$ represents a $K \times 1$ vector of time-invariant regression coefficients. The terms associated with the $TK$ TVPs $\bm \gamma = (\bm \gamma'_1, \dots, \bm \gamma'_T)'$ are put in $\bm Z$ which is a $T \times (TK)$ block-diagonal matrix defined by:
\begin{equation*}
\bm Z = 
\begin{pmatrix} 
  \bm x_1'  & \bm 0_{K \times 1}'   & \dots  & \bm 0_{K \times 1}' \\
 \bm 0_{K \times 1}'    & \bm x_2' & \dots  & \bm 0_{K \times 1}' \\ 
  \vdots    & \vdots   & \ddots & \vdots \\
  \bm 0_{K \times 1}'    & \bm 0_{K \times 1}'   & \dots  & \bm x_T'
\end{pmatrix}.
\end{equation*}
For simplicity of exposition, for now we assume homoskedastic errors and, thus, $\varepsilon \sim \mathcal{N}(\bm 0, \sigma_\varepsilon^2 \bm I_T)$. The addition of stochastic volatility is discussed in Sub-section \ref{sec: SVest}.

Note that the regression coefficients are now divided into $\bm \beta$, which are conventional constant coefficients, and $\bm \gamma$ which are the TVPs. At this stage, our treatment assumes the latter to be completely unrestricted. For instance, we are not restricting them to follow random walk processes as is commonly done with TVP regressions. This specification implies that the coefficient on each explanatory variable is $\beta_{it} = \beta_i + \gamma_{it}$, with $\gamma_{it}$ denoting the $i^{th}$ element of $\bm \gamma_t$.

It is also worth noting that the methods we develop can be easily extended to the TVP- VAR if it is written in equation-by-equation form \citep[see, e.g.,][]{kastnerhuber2017,ccm2019, kkp2019, hko2019}. In such a case, the TVP-VAR with $N$ variables turns into $N$ TVP regressions. Appendix \ref{sec: appVAR} shows how this can be done. In the subsequent discussion, we simplify notation by assuming that $N=1$.  For the empirical application, all hyperparameters are specified symmetrically across equations so that regression coefficients and error covariances are treated in the same way in each equation.

\subsection{The Prior}
\label{prior}

In the case where $K$ is large, it is empirically probable that $\bm \beta$ is much more important (i.e. having some non-zero elements) than $\bm \gamma$, most of whose elements are likely to be zero. This consideration motivates both our prior choice and our choice of devoting most of the computational effort to the former rather than the latter. In particular, we obtain an approximation of the posterior for $\bm \gamma$ using computationally-fast techniques, then sample from the posterior of $\bm \beta$ using conventional MCMC methods conditional on this approximation. Posterior analysis is discussed in the following sub-section, in this sub-section, we discuss the prior. 

The methods developed in this paper will work for any prior which has a Gaussian hierarchical structure. So, for instance, any of the global-local shrinkage priors which involve Gaussianity at the first stage of the prior hierarchy can be used. See, for instance, \cite{bppd2015} for a discussion of some global-local shrinkage priors and their properties. 

For constant coefficients, $\bm \beta$, we use the popular Normal-Gamma (NG) shrinkage prior of \cite{griffin2010}:
\begin{equation*}
p(\bm \beta) = \prod_{j=1}^K \mathcal{N}(\beta_j | 0, \tau_j^2),~ \tau_j^2 \sim \mathcal{G}(\vartheta, \vartheta \lambda/2), \lambda \sim \mathcal{G}(d_0, d_1),
\end{equation*}
with $\vartheta$ being a hyperparameter that controls the excess kurtosis of the marginal prior (obtained by integrating out the local scaling parameters $\tau_j$). $\lambda$ is a global shrinkage parameter that pushes all elements in $\bm \beta$ to zero. It follows a Gamma distribution with parameters $d_0$ and $d_1$ a priori. In the empirical work, we follow much of the literature and set $\vartheta=0.1$ and $d_0=d_1=0.01$. This choice places significant prior mass on zero but allows for heavy tails.

For $\bm \gamma$, we rely on two commonly used shrinkage priors. The first one is a Spike \& Slab (S\&S) prior implemented as in \cite{mitchell1988bayesian}. This is a mixture of two distributions, one (the Spike) with a point mass at zero and the other (the Slab) a zero mean Gaussian prior with variance equal to $\psi$. Formally, this prior is given by:
\begin{equation*}
p(\bm \gamma) = \prod_{j=1}^{TK} q \mathcal{N}(0, \psi) + (1-q) \delta (0).
\end{equation*}
$q$ denotes the prior probability of observing a non-zero element in $\bm \gamma$, i.e. $q=Prob(\gamma_j \neq 0)$ and $\delta (0)$ is a point mass at zero. In all applications, we set $q$ equal to $0.5$. For the prior hyperparameter $\psi$, we consider a range of values as discussed in the empirical section of this paper.  Since, as noted above, our data is standardized, the assumption of a single prior hyperparameter, $\psi$, is a reasonable one.

The second prior is the sparse Bayesian learning (SBL) specification originally proposed in \cite{tipping2001sparse} and subsequently adopted in \cite{zou2016computationally} and \cite{korobilis2019high}. This prior assumes that
\begin{align*}
p(\bm \gamma) = \prod_{j=1}^{TK} \mathcal{N}(0, \psi_j),\quad \psi^{-1}_j \sim \mathcal{G}(a_\psi, b_\psi),
\end{align*}
with $a_\psi =1$ and $b_\psi= 10^{-6}$ denoting prior hyperparameters with the specific choices motivated in \cite{fang2016two}. This conditionally Gaussian prior has been shown to possess excellent shrinkage properties in many contexts \citep[see, e.g.,][]{korobilis2019high}. 

To explain these prior choices, remember that the (low dimensional) constant coefficients are more likely to be non-zero than the (high dimensional) TVPs. 
For the constant coefficients, we use a popular global-local shrinkage prior which has been found to work well in many empirical applications. It allows for unimportant coefficients to be shrunk to be close to zero (not precisely zero). The Spike \& Slab prior, by contrast, allows for unimportant coefficients to be precisely zero. \cite{hko2019} demonstrates the importance of shrinking coefficients to be precisely zero in high dimensional cases. That is, if the number of coefficients that equal zero  is large, then a global-local shrinkage prior can successfully shrink them to be near to zero, but not exactly zero. The small amount of estimation error in each of a large number of coefficients can accumulate and cause forecast performance to deteriorate. This is why we use the Spike \& Slab prior for the high dimensional vector of coefficients, $\bm \gamma$.  Bayesians using MCMC methods often avoid the use of the Spike \& Slab prior since it can suffer from poor MCMC mixing. However, since we are not using MCMC methods for $\bm \gamma$ this difficulty of the Spike \& Slab prior is not relevant for us.  

As opposed to the Spike \& Slab prior, which only allows for discriminating between the case that a coefficient is time-varying or constant for a given point in time, the SBL prior  allows for cases in between and thus might capture situations where the coefficient is constant, mildly time-varying or strongly time-varying at certain points in time.

Finally, we use a weakly informative inverted Gamma prior on $\sigma_\varepsilon^2 \sim \mathcal{G}^{-1}(0.01, 0.01)$. 

\subsection{Posterior Analysis Using Integrated Gaussian Rotations}\label{sec: IRGA}
Following \citet{boom2019approximating}, we use IRGAs to separately model the TVPs and the constant coefficients. This approach exploits properties of the QR decomposition of  $\bm X$. The QR decomposition says we can write the $T \times K$ matrix $\bm X = \bm Q \bm R$ where $\bm Q$ is a $T \times T$ orthogonal matrix and $\bm R$ is a $T \times K$ upper triangular matrix with the final $T-K$ rows being zero. We partition $\bm Q = (\bm Q_{1}, \bm Q_{2})$ where $\bm Q_{1}$ is $T \times K$ and $\bm Q_{2}$ is $T \times (T-K)$. A key property of the QR decomposition, which we will rely on below, is $\bm Q_{2}' \bm X = \bm 0_{(T-K) \times K}$. Note also that these results for the QR decomposition assume $T \ge K$. We will discuss the $T<K$ case in a subsequent sub-section.  

The TVP regression model given in (\ref{eq: regmodel}) implies the following distribution for $\bm y$:  
\begin{equation}
\bm y \sim \mathcal{N}( \bm X \bm \beta +  \bm Z \bm \gamma, \sigma_\varepsilon^2 \bm I_K).
 \label{likey}
\end{equation}
Instead of working directly with the likelihood defined by (\ref{likey}), what the IRGA approach does is exploit the fact that Gaussian distributions are invariant to rotations. It is based on the following 
two distributions:
\begin{equation}
\bm Q_{1}' \bm y \sim \mathcal{N}(\bm Q_{1}' \bm X \bm \beta + \bm Q_{1}' \bm Z \bm \gamma, \sigma_\varepsilon^2 \bm I_K), \label{likebeta}
\end{equation}
and
\begin{equation}
\bm Q_{2}' \bm y \sim \mathcal{N}(\bm Q_{2}' \bm Z \bm \gamma, \sigma_\varepsilon^2 \bm I_{T-K}).
\label{likeeta}
\end{equation}
Note that, since $\bm Q_{2}' \bm X = \bm 0_{(T-K) \times K}$, $\bm \beta$ does not appear in the second (high dimensional) relationship and, thus, if $\bm \gamma$ were known, posterior inference for it can be based on the first (low dimensional) relationship. This form motivates a two stage approach where a computationally fast approximation is first done to produce an approximate posterior for $\bm \gamma$ (as well as any hyperparameters associated with the prior on $\bm \gamma$) based on (\ref{likeeta}) and subsequently posterior inference on $\bm \beta$ conditional on this is done using a likelihood based on (\ref{likebeta}). 

To be precise, the first stage involves obtaining a Gaussian approximation to the posterior of $\bm \gamma$ using a regression with $\bm Q_{2}' \bm y$ as the dependent variable:
\begin{align*}
\hat{p}(\bm \gamma | \bm Q_{2}' \bm y, \bm Q_2' \bm Z) \sim \mathcal{N}(\bm \mu_\gamma, \bm V_\gamma).
\end{align*}
The mean and variance of $\hat{p}(\bm \gamma | \bm Q_{2}' \bm y, \bm Q_2' \bm Z)$ can be obtained using any sort of Gaussian approximation. In this paper we use VAMP. Message passing algorithms have been used successfully in TVP regressions in \cite{korobilis2019high}. Note that these algorithms are particularly likely to be accurate when the correlations between explanatory variables are low. Since the form of our $\bm Z$ matrix implies blocks of variables for different time periods are uncorrelated with one another, message passing algorithms seem appropriate for our case. VAMP, originally proposed in \cite{rangan2019vector}, differs from standard approximate message passing (AMP) algorithms by using a particularly simple (i.e. scalar-valued) state evolution equation to update the posterior for $\bm \gamma$. More importantly, VAMP is more widely applicable relative to AMP in the sense that VAMP can be applied to a larger class of design matrices \citep{rangan2019vector}. For a detailed implementation of the VAMP algorithm that includes information on how we estimate $\bm \gamma$ and the posterior inclusion probabilities, see the technical appendix of \cite{boom2019approximating}. 

To update $\psi_j$ under the SBL prior, we follow \cite{zou2016computationally, korobilis2019high} and update  $\psi_j$ within VAMP as follows:
\begin{equation*}
\psi^{(n+1)}_j = \frac{2a_\psi - 1}{\left(\hat{\gamma}^{(n)}_j \right)^2+ 2 b_\psi}.
\end{equation*}
Here, we let $\hat{\gamma}^{(n)}_j$ denote the estimate of the $j^{th}$ element of $\bm \gamma$ at iteration $n$ of the VAMP algorithm.

Plugging in the values of $\bm \mu_\gamma$ and $\bm V_\gamma$ produced by the VAMP algorithm in the first stage, the approximate posterior of $\bm \beta$ in the second stage is based on:
\begin{align*}
\hat{p}(\bm \beta |  \bm y) \propto p(\bm \beta) \times \mathcal{N}\left(\bm Q_{1}' \bm y | \bm Q_{1}' \bm X \bm \beta + \bm Q_{1}'  \bm Z \bm{\mu}_{\gamma}, \sigma_\varepsilon^2 \bm I_K + (\bm Q_1' \bm Z) \bm V_\gamma (\bm Q_1' \bm Z)'  \right),
\end{align*}
where  $\bm \Omega_{Q_1} = \sigma_\varepsilon^2 \bm I_K + (\bm Q_1' \bm Z) \bm V_\gamma (\bm Q_1' \bm Z)'$ is a full error variance-covariance matrix with its lower Cholesky factor given by $\bm P_{Q_1}$ and $p(\bm \beta)$ is the prior.  As noted above, this prior can take any form, but if it is Gaussian (or of a hierarchical form with the first level being Gaussian), the Gaussian prior and Gaussian likelihood will combine to produce a Gaussian posterior (conditional on the parameters in the hierarchical prior). This Gaussianity property combined with the fact that $\bm \beta$ is only $K$ dimensional implies computation is fast. Textbook results for the linear regression model with dependent variable $\tilde{\bm y} = \bm  P_{Q_1}^{-1}(\bm Q_{1}' \bm y - \bm Q'_{1} \bm Z \bm \mu_\gamma)$ and explanatory variables $\tilde{\bm X} = \bm P_{Q_1}^{-1}(\bm Q_{1}' \bm X)$ provide the arguments in this posterior (or conditional posterior if, as we do, a hierarchical shrinkage prior is used). More precisely, this posterior is given by:\footnote{It is noteworthy that the error variance-covariance is a full matrix but, conditional on multiplying from the left with the inverse of its lower Cholesky factor $\bm P_{Q_1}^{-1}$, yields a linear regression model with standard normally distributed errors. }
\begin{equation}
\bm \beta | \bm y \sim \mathcal{N}(\bm \mu_\beta, \bm {V}_\beta) \label{eq:postBeta}
\end{equation}
with posterior moments:
\begin{align*}
\bm V_\beta &= (\tilde{\bm X}' \tilde{\bm X} + \bm \Lambda^{-1})^{-1}\\
\bm \mu_\beta &= \bm V_\beta (\tilde{\bm X}' \tilde{\bm y}).
\end{align*}
We let $\bm \Lambda = \text{diag}(\tau_1^2, \dots, \tau_K^2)$ denote a diagonal prior variance-covariance matrix.

As noted above, we use the familiar shrinkage prior of \cite{griffin2010}. With this choice, the MCMC algorithm involves adding on to textbook MCMC for the Gaussian linear regression model with Gaussian prior, methods for drawing the shrinkage parameters, $\tau_j^2$, for $j=1,\ldots,K$ and $\lambda$, which appear in the NG prior. $\tau_j^2$ is simulated from its generalized inverted Gaussian (GIG) full conditional posterior distribution:\footnote{We simulate from the GIG distribution using the R package GIGrvg and the corresponding parameterization.}
\begin{equation}
\tau_j^2| \alpha_j, \lambda \sim \mathcal{GIG}(\vartheta-1/2, \alpha_j^2, \lambda \vartheta)\label{eq: posttau}
\end{equation} 
 and $\lambda$ has a Gamma-distributed conditional posterior distribution:
 \begin{equation}
 \lambda | \tau_1^2, \dots, \tau_K^2 \sim \mathcal{G}\left(d_0 + K \vartheta, d_1 +  \vartheta \frac{\sum_{j=1}^K \tau^2_j}{2}\right).\label{eq: postlambda}
 \end{equation}

In principle, the error variance $\sigma^2_\varepsilon$ can be estimated using either (\ref{likebeta}) or (\ref{likeeta});  \cite{boom2019approximating} use the second stage regression in  (\ref{likeeta}) to estimate $\sigma^2_\varepsilon$. This implies that only   $(T-K)$ observations are used to inform the posterior estimates of $\sigma^2_\varepsilon$. However, in the case where $T > K$ (which is often the case for monthly data), this is a sensible strategy since we would expect that the second-stage regression contains significant information on the error variance. 

\subsection{Allowing for Stochastic Volatility} \label{sec: SVest}
In macroeconomic forecasting applications, allowing for conditional heteroscedasticity has been shown to be of great importance to produce accurate point and density forecasts, see e.g. \cite{clark2011}.  Introducing SV in our model, however, is not straightforward. The fact that we split the likelihood in two parts means that we are not able to estimate a time-varying variance for all $t$ without additional complications. 

As a simple solution, we modify (\ref{eq: regmodel}) as follows:
\begin{align}
\bm y &= \bm X \bm \beta + \bm Z \bm \gamma + \bm \eta + \bm \varepsilon, \nonumber \\
\eta_ t &= e^{h_t/2} \varepsilon_{\eta, t} \text{ for } t=1,\dots,T. \label{eq: etaSV}
\end{align}
Here, $\bm \eta = (\eta_1, \dots, \eta_T)'$ is a process that features SV and $h_t$ denotes a log-volatility process while $\varepsilon_{\eta, t} \sim \mathcal{N}(0,1)$. This specification implies that the variance of the sum of $\bm \eta$ and $\bm \varepsilon$ is $\sigma_t^2 = e^{h_t}+\sigma^2_\varepsilon$ and that neither $h_t$ nor $\sigma_\varepsilon^2$ are separately identified. 

Estimating $\bm \eta$  and the full history of $h_t$ is more complicated. Our proposed approach, however, is perfectly suited for handling such a situation. To be more precise, we approximate $\bm \eta$ and $\{h_t\}_{t=1}^T$ alongside $\bm \gamma$ exploiting equation (\ref{likeeta}) while running VAMP.  This is simply achieved by noting that $\bm \eta$ is a time-varying intercept term that features SV. Estimating $h_t$ is, unfortunately, more difficult.   In this paper, we adopt the SV estimator proposed in  \cite{korobilis2019high}. Specifically, we square and take logs of (\ref{eq: etaSV}) to obtain:
\begin{equation*}
\tilde{\eta}_t = h_t + v_t, 
\end{equation*}
with $\tilde{\eta}_t = \log \eta^2_t$ and $v_t= \log \varepsilon^2_{\eta, t}$ denotes an error term that follows a log $\chi^2$ distribution with one degree of freedom. \cite{kim1998stochastic} propose approximating the distribution of $v_t$ using a seven component mixture of Gaussians. This renders the observation equation conditionally Gaussian:
\begin{equation*}
\tilde{\eta}_t = h_t + u_i, \quad u_i \sim \mathcal{N}(\mu_{v, i},  {\sigma}^2_{v, i}),
\end{equation*}
for $i=1,\dots,7$. Here, we let $\mu_{v, i}$  and ${\sigma}^2_{v, i}$ denote component-specific means and variances, respectively. Let $\pi_i$ denote the corresponding mixture weights. As a point estimator for $h_t$, we then use
\begin{equation*}
\hat{h}_t = \sum_{i=1}^7 \pi_i (\tilde{\eta}_t - \mu_{v, i}).
\end{equation*}
The precise values for $\mu_{v, i}, {\sigma}^2_{v, i}$ and $\pi_i$ can be found in  \cite{kim1998stochastic}, Table 4.  \cite{korobilis2019high}  shows that this volatility estimate possesses excellent empirical properties and also closely matches estimates obtained by using MCMC-based techniques but tend to be less persistent. This is because $\hat{h}_t$ does not depend on $\hat{h}_{t-1}$.

\subsection{Posterior simulation}
To estimate the posterior distribution of the coefficients of the model as well as the latent states, we use the following algorithm:\footnote{We assume throughout that everything has been initialized by using maximum likelihood estimates (if available) or by starting from the prior mean.}
\begin{steps}
\item Approximate $\bm \gamma$ (and the corresponding hyperparameters associated with the prior)$, \sigma^2_\varepsilon$ and $\{ h_t \}_{t=1}^T$ by exploiting (\ref{likeeta}) and the methods outlined in Sub-sections \ref{sec: IRGA} and \ref{sec: SVest}.

\item Conditional on having approximated the TVPs und log-volatilities,  obtain a draw from  $p(\bm \beta|\bm y)$ using (\ref{eq:postBeta}).

\item Draw $\{\tau_j^2\}_{j=1}^K$ from the GIG distribution in (\ref{eq: posttau}).

\item Finally, simulate $\lambda^2$ from a Gamma posterior detailed in (\ref{eq: postlambda}).
\end{steps}
In all empirical applications, we repeat Steps (2) to (4) $15,000$ times and discard the first $10,000$ draws as burn-in. Since Step 1 is fast, estimating a TVP regression model using our IRGA approach takes approximately as long as estimating a constant parameter specification, significantly reducing computation time. Further details on computation times for the different applications can be found in Sections \ref{sec: artData} and \ref{sec: empApp}.

\subsection{Methods when $T<K$}
\label{sec: TltK}
An important limitation of the IRGA methods described in Sub-section \ref{sec: IRGA} is that they require $T \ge K$ to work.  But in multivariate time series models such as TVP-VARs we often encounter the situation that $T<K$. That is, if the number of endogenous variables ($N$) and the number of lags ($P$) is large, the number of explanatory variables in each equation (i.e. $NP$ plus any exogenous regressors)  can be enormous. Given the number of observations typically associated with macroeconomic quarterly or monthly data sets, many relevant empirical settings have $T<K$.

In such a case, we would recommend that the best procedure is to reformulate the problem so that $T \ge K$. Remember the basic idea of our approach is that there are $K$ parameters which are important and others which can be treated as nuisance parameters. When $T \ge K$ it is logical to class the constant coefficients as the important parameters and the TVPs as nuisance parameters.  However, it is also possible to classify some of the less important constant coefficients as nuisance parameters.  For instance, in a macroeconomic TVP-VAR there will typically be some variables of primary interest either for forecasting or for economic structural analysis  (e.g. in our empirical section we use interest rates, output growth and inflation as the variables of primary interest). Constant coefficients on the lagged values of these primary variables can be estimated using MCMC methods, with all remaining parameters (i.e. constant coefficients on lagged values of other variables as well as TVPs) being treated as nuisance parameters. 
 
Alternatively, in macroeconomic models such as VARs it is common to find that lower order lags of endogenous variables are more important than higher order lags. Thus, in some applications it might make sense to choose a small number of lags, $P_1$, and treat coefficients on all lagged endogenous variables up to $P_1$ as being of primary interest, with coefficients on the remaining $P-P_1$ lags being put into the category of nuisance parameters.  

In our empirical work, we consider two approaches using a TVP-VAR which assumes that $\bm y_t$ is an $N$-dimensional vector of endogenous variables (see Appendix \ref{sec: appVAR}). One is the $T \ge K$ case outlined above which treats only the TVPs as nuisance parameters. The second treats the TVPs and all cross-lags greater than one as nuisance parameters. As we will show in Sub-section \ref{sec: forecasting}, this only slightly decreases predictive accuracy but allows for estimating even larger models while still maintaining fast computation times.

\section{Artificial Data}\label{sec: artData}
In this section, we assess how our approach performs when applied to synthetic data. To this end, we use a data generating process (DGP) based on (\ref{eq: regmodel}). Each DGP takes a draw of constant coefficients assuming  $\bm \beta \sim \mathcal{N}(0, \bm I_K)$ and the elements of $\bm \gamma$ are drawn as:
\begin{align*}
\gamma_{it} \sim \mathcal{N}\left(0, s_t (0.5)^2\right), \quad
s_t = \begin{cases} 1 \text{ with probability $\underline{p}$}\\
0 \text{ with probability $1-\underline{p}$}, \end{cases}\quad
\bm x_t &\sim \mathcal{N}(\bm 0_K, \bm I_K),
\end{align*}
for $t=1,\dots, 500$ and with $\underline{p}$ denoting the probability of observing a parameter change. We set $\sigma_\varepsilon = 0.1$. We consider a range of values for $\underline{p}$. If we set $\underline{p}$ equal to zero we obtain a constant parameter specification while setting $\underline{p}=1$ yields a standard TVP model. To investigate how our approach performs across a wide variety of DGPs, we choose $K \in \{5, 10, 15, 25\}$ and $\underline{p} = \{0, 0.25, 0.5, 0.75, 1\}$ and obtain $100$ realizations of the DGP. To illustrate that approximating the TVPs using VAMP allows us to estimate the time-invariant part of the model efficiently, we benchmark our approach (henceforth labeled IRGA-TVP S\&S for the Spike \& Slab prior and IRGA-TVP SBL for the SBL prior) to a TVP regression model estimated using standard MCMC techniques and shrinkage priors \citep[see, e.g., ][]{bitto_fs, hko2019} and a random walk state equation.

Note that our approach is an approximate one whereas our benchmark is more exact (if sufficient MCMC draws are taken). This represents an advantage for the benchmark approach. However, our approach allows for different types of TVPs. In addition, IRGA-TVP S\&S allows for them to be shrunk to be precisely zero, whereas the benchmark approach does not since it assumes random walk behavior. We expect this, in many empirical contexts, to be an advantage for our approach. That is, in practice, it is likely to be the case that many or most of the TVPs are zero, particularly when the number of explanatory variables is large.

To see how these considerations play out in artificial data, consider Table \ref{tab: sim} which displays the mean absolute errors (MAEs) for $\bm \beta$  of the IRGA approach  relative to the standard TVP regression. In general, and for both priors considered, we see numbers near one, indicating that both approaches are yielding similar estimates, despite the fact that our IRGA approaches are much more computationally efficient. Using  IRGA in combination with the Spike \& Slab prior yields slightly worse estimates  than the benchmark when $\underline{p}=1$, but performs appreciably better in all other cases. But the $\underline{p}=1$ case allows time variation in every coefficient and in every time period. It is not surprising that the benchmark model, which allows for this high degree of time variation, performs well here. And it is worth noting that, other than for the case where $K=5$ to which we will return later, our algorithm is doing roughly as well as the benchmark even with $\underline{p}=1$. Using the SBL prior in this situation yields estimates that are very close to the ones of the benchmark, highlighting the increased flexibility by allowing for a varying degree of time variation over the estimation period. 

When we move to DGPs with less time variation and, thus, smaller values for $\underline{p}$, our methods do appreciably better relative to the benchmark (but to a slightly lesser extent if the SBL prior is used). Our algorithm does particularly well for intermediate values of $\underline{p}$, but even in the constant coefficient case, $\underline{p}=0$, we are still beating the benchmark under both priors adopted. Note that the shrinkage prior in the benchmark approach allows for separate shrinkage on constant coefficients and TVPs. Thus, it allows for any coefficient to be constant (over the entire sample) or time-varying (over the entire sample). But, unlike our approach, it does not allow for coefficients to be time-varying in some periods or not others. This is likely why our methods do particularly well for values in the region of $\underline{p}=0.5$. Artificial data sets generated from such a value will have many periods where coefficients are constant and others where they are varying. DGPs with the extreme choices of $\underline{p}=0$ or $\underline{p}=1$ will not have this property and lead to relatively better performance of the benchmark. This also explains the small differences between the Spike \& Slab and the SBL priors, with the latter having a slight disadvantage since they are not capable of pushing the TVPs exactly to zero.

Next we consider the impact of $K$ on the relative MAEs. When $K=5$ and $\underline{p}=1$ we have a small TVP regression with lots of parameter change. In such a case, the benchmark does well. However, even with $\underline{p}=0$ and larger values for $K$, our methods improve relative to the benchmark. Our best relative performance occurs for the largest model with $K=25$ and an intermediate amount of parameter change $\underline{p}=0.75$. This largely reflects the fact that when using a Spike \& Slab prior for the TVPs, we allow them to be shrunk precisely to zero in periods when there is no parameter change. This contrasts with the SBL prior and the benchmark approach which only allow for them to be shrunk to be near zero. \cite{hko2019} show that, in models with many parameters, shrinkage to precisely zero is typically useful in achieving forecast accuracy. Other forms of shrinkage allow for small amounts of estimation error to accrue over many parameters, which can be detrimental for forecast accuracy. This effect can also be clearly seen in our results in terms of the impact on estimation accuracy for the constant coefficients. 

\begin{table}[ht]
\centering
\begin{tabular}{lrrrrr}
  \toprule
  & \multicolumn{5}{c}{Amount of time-variation $\underline{p}$}\\
  $K$   & $1$ & $0.75$ & $0.5$ & $0.25$ & $0$ \\ 
    \multicolumn{6}{c}{S\&S prior}\\\midrule
5 & 1.616 & 0.681 & 0.771 & 0.773 & 0.926 \\ 
  10  & 1.064 & 0.754 & 0.657 & 0.645 & 0.773 \\ 
  15 & 1.049 & 0.629 & 0.654 & 0.769 & 0.727 \\ 
  20 & 1.102 & 0.667 & 0.803 & 0.715 & 0.744 \\ 
  25 & 1.179 & 0.583 & 0.798 & 0.685 & 0.740 \\
    \multicolumn{6}{c}{SBL prior} \\\midrule
  5 & 1.043 & 1.032 & 0.999 & 0.892 & 0.931 \\ 
  10 & 1.029 & 1.004 & 0.922 & 0.786 & 0.929 \\ 
  15 & 0.968 & 0.834 & 0.783 & 1.052 & 0.739 \\ 
  20 & 0.995 & 0.799 & 0.918 & 0.862 & 0.799 \\ 
  25 & 0.969 & 0.744 & 0.763 & 0.729 & 0.687 \\ 
   \bottomrule
\end{tabular}
\caption{Mean absolute error ratios between a TVP model estimated using our IRGA approach and a constant parameter model. All results are based on $100$ realizations from the DGP.} \label{tab: sim}
\end{table}

\begin{figure}[h!]
\centering
\includegraphics[scale=.6]{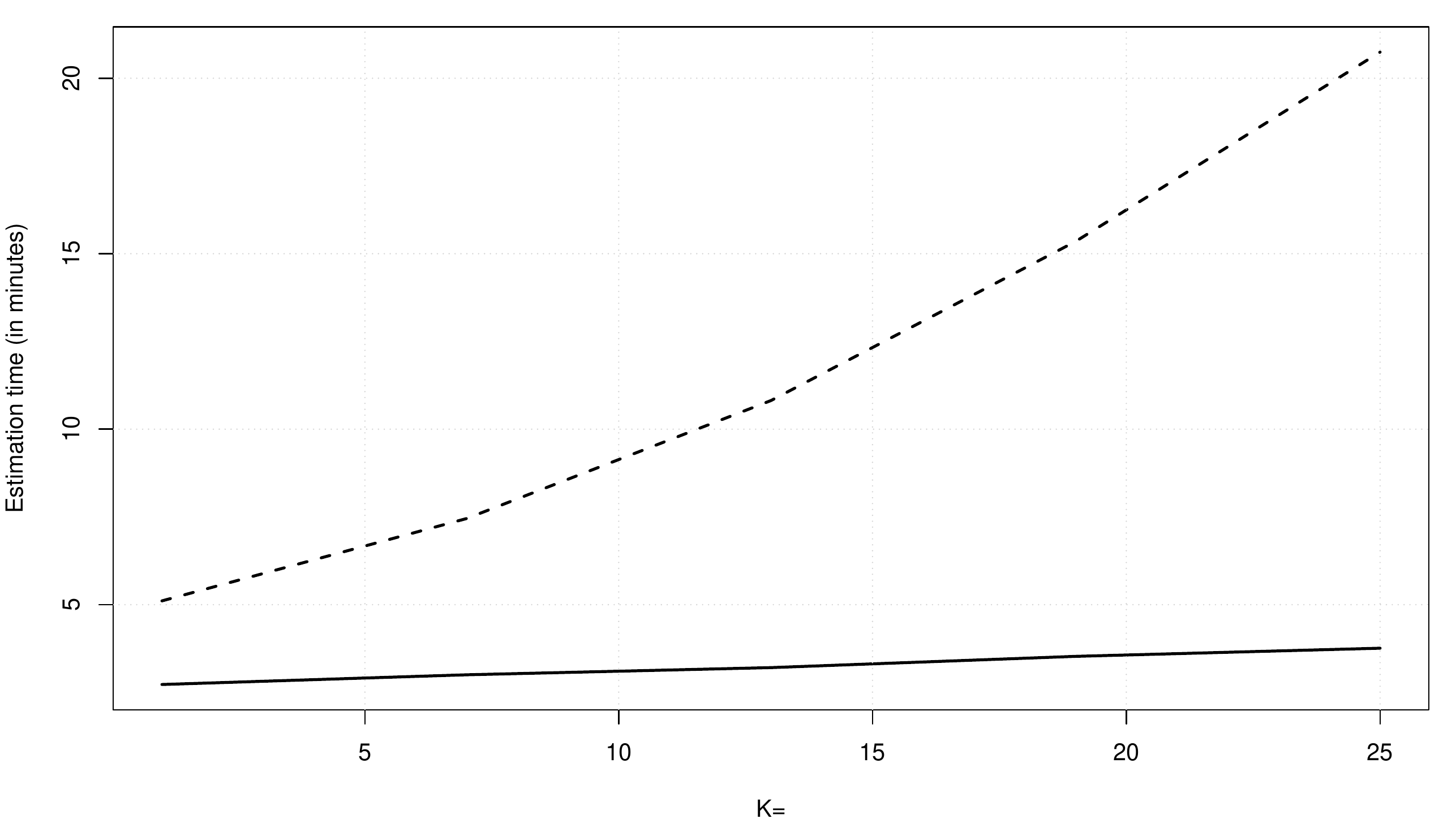}
\caption{Time necessary for producing $15,000$ draws from the joint posterior: IRGA - TVP (solid black) and  TVP (dashed black).}\label{fig: estim_time}
\end{figure}

We next turn to the issue of computation time. Figure \ref{fig: estim_time} depicts the time necessary (in minutes) to generate $15,000$ draws from the joint posterior for the benchmark TVP regression (in dashed black) and our IRGA-TVP approach (in solid black). The key point is that our IRGA methods are scaleable and can potentially handle hundreds of explanatory variables, whereas the benchmark approach cannot. To be precise, estimation times increase linearly at a very slow rate with $K$ under the IRGA approach, with overall time ranging  from 2.7 to around 3.7 minutes.  By contrast, estimating a standard TVP regression model using the algorithm proposed in \cite{carter1994gibbs} is  much more demanding, with estimation times for this model increasing quite rapidly at a nonlinear rate. 

Summing up, we have shown that the estimates for $\bm \beta$ obtained using our IRGA approach are precise and close to the ones obtained from estimating a standard TVP regression. In addition, approximating the TVPs leads to vastly improved estimation times that only increase slightly with the number of predictors.

\section{Empirical Work Using US Macroeconomic Data Set}\label{sec: empApp}
In this section, we carry out a forecasting exercise and structural economic exercise using a large TVP-VAR involving up to $N=78$ variables (see Appendix \ref{app:dataresults}). 
 
\subsection{Data}
We use the popular FRED-MD macroeconomic data set, see \citet{mccracken2016fredmd}. We obtain monthly data from 1985:01 to 2019:08 on $78$ variables. The exact variables (and the way each has been transformed to stationarity) are listed in Table \ref{data_table} in Appendix \ref{app:dataresults}. We use measures of inflation (labelled CPIAUCSL), interest rates (FEDFUNDS) and the unemployment rate (UNRATE) as our three variables of primary interest. The remaining variables are only useful insofar as they improve forecasts or affect structural impulse responses to the primary variables.

\subsection{Forecasting Exercise}\label{sec: forecasting}
We carry out a forecasting exercise using a huge data set involving $78$ variables. The results are based on four different versions of our IRGA approach involving two priors (the SBL and S\&S priors) and the two methods of implementation outlined at the end of Sub-section \ref{sec: TltK}. These are labelled "TVPs" and "TVPs \& lag" below. Prior hyperparameter choice is discussed in Sub-section \ref{prior}. For the S\&S prior we fix one of the key prior hyperparameters by setting $\psi=0.001$. A prior sensitivity analysis with respect to this hyperparameter is done in Appendix \ref{app:dataresults}. The lag length is set equal to two.

A natural benchmark for a dataset of this size is a Bayesian VAR with a hierarchical Minnesota prior \citep{litterman1986forecasting} that entails cross-variable and cross-equation shrinkage. The three hyperparameters governing the degree of shrinkage on the VAR coefficients are integrated out using weakly informative Gamma priors (with all  hyperparameters of the Gamma distributions set equal to $0.01$) and a random walk Metropolis step, similar to \cite{giannone2015prior}. For the error covariances and the intercept term, we use normally distributed priors with zero mean and variance ten. The model is estimated using equation-by-equation estimation. 

As a a low dimensional benchmark, we include a three equation TVP-VAR with SV using just the focus variables. This is a standard TVP model which assumes the TVPs follow random walks. It is estimated using MCMC methods on an equation-by-equation basis using a non-centered specification with Gaussian prior with zero mean and variance ten on all parameters.

All benchmark models use a standard stochastic volatility specification that assumes an AR(1) evolution of the log-volatilities and is estimated using the algorithm put forward in \cite{kastner2014ancillarity}.

We investigate the performance of our methods for the three primary variables of interest: inflation, unemployment and the interest rate. We use root mean squared error (RMSE) and average log-predictive likelihoods (LPLs) to evaluate forecast performance. We use a forecast evaluation period beginning in 2005:01 and running to the end of the sample for forecast horizons of one month and one year.

Before discussing forecast performance, it is worthwhile to briefly consider computation time. Table \ref{tab:esttimefcst} presents  average estimation times taken by the forecast exercise over the holdout sample. For the IRGA specifications approximating solely the TVPs, the sampling algorithm takes roughly $150$ minutes. When we approximate second-order cross-variable lags alongside the TVPs, estimation is even faster. It takes about $125$ minutes.  Note that the IRGA specifications allow for parallelization of the algorithm (for reference, we only used one CPU core). Estimation using multiple cores speeds up estimation by a factor of the number of available CPUs (e.g. eight cores would allow to estimate the models eight times as fast, and the $78$ equation IRGA could be estimated in around $20$ minutes). 

For reference, the hierarchical Minnesota BVAR takes $270$ minutes to estimate, while the small TVP-VAR takes roughly four minutes (however, as suggested above, scales badly in $T$ and $K$). One would expect that estimation times of the constant parameter VAR and one of the IRGA models are comparable, which is clearly not the case here. This  difference in runtimes arises from the fact that the BVAR features a standard SV specification and this implies that, if equation-by-equation estimation is used, we need to normalize $\bm y$ and $\bm X$ by dividing through the time-varying error standard deviation and repeatedly compute a cross product involving $\bm X$ during MCMC sampling.


\begin{table}[h!]
\caption{Average estimation time over the holdout.}\label{tab:esttimefcst}\vspace*{-1.3em}
\begin{center}
\begin{threeparttable}
\footnotesize
\begin{tabular}{lcccc}
\toprule
\textbf{Model} & \textbf{IRGA SBL (TVPs)} & \textbf{IRGA SBL (TVPs \& lag)} & \textbf{Minnesota} & \textbf{TVP-VAR}\\ 
\midrule
Runtime & $153.8$ & $125.6$ & $270.0$ & $3.5$\\
\bottomrule
\end{tabular}
\begin{tablenotes}[para,flushleft]
\scriptsize{\textit{Notes}: Runtimes in minutes.}
\end{tablenotes}
\end{threeparttable}
\end{center}
\end{table}

Results of our forecasting exercise are given in Table \ref{tab:fcst}. RMSEs shed light on point forecasting performance and indicate our methods are forecasting roughly as well as the benchmark Bayesian VAR. With one exception, the same can be said for the three variable TVP-VAR. The one exception is for long-run forecasting ($h=12$) of the interest rate where the TVP-VAR is forecasting the interest rate very poorly. In contrast, the IRGA approaches (in particular, the S\&S variants) are forecasting very well. This suggests that, for the interest rate, there is little evidence of time-variation in parameters. In this case, the S\&S prior is doing a very good job of shrinking excess time-variation to be zero in a way that the TVP-VAR does not. 

LPLs evaluate the qualities of entire predictive densities produced by the different approaches. Note that all of the LPLs for the IRGA approach are positive, indicating that they are beating the benchmark Bayesian VAR. These gains are often pronounced, strongly highlighting the benefits of combining flexible models with large information sets using the IRGA framework. We stress that all of our models, including the Bayesian VAR, have stochastic volatility. Our findings are suggesting that, even in large VARs, there is some improvement in predictive densities to be found by allowing for time variation in both VAR coefficients and error covariance matrices. Note also that the low-dimensional TVP-VAR is producing predictive densities which are roughly as good as our high-dimensional IRGA approaches. This reinforces the idea that there may be a trade-off between TVPs and VAR dimension. In small data sets, TVPs may be particularly important as they may reduce the effect of omitted variables bias whereas in large dimensions, this bias is substantially reduced by controlling for observed heterogeneity through the inclusion of additional covariates. 

\begin{figure}[ht]
\includegraphics[width=\textwidth]{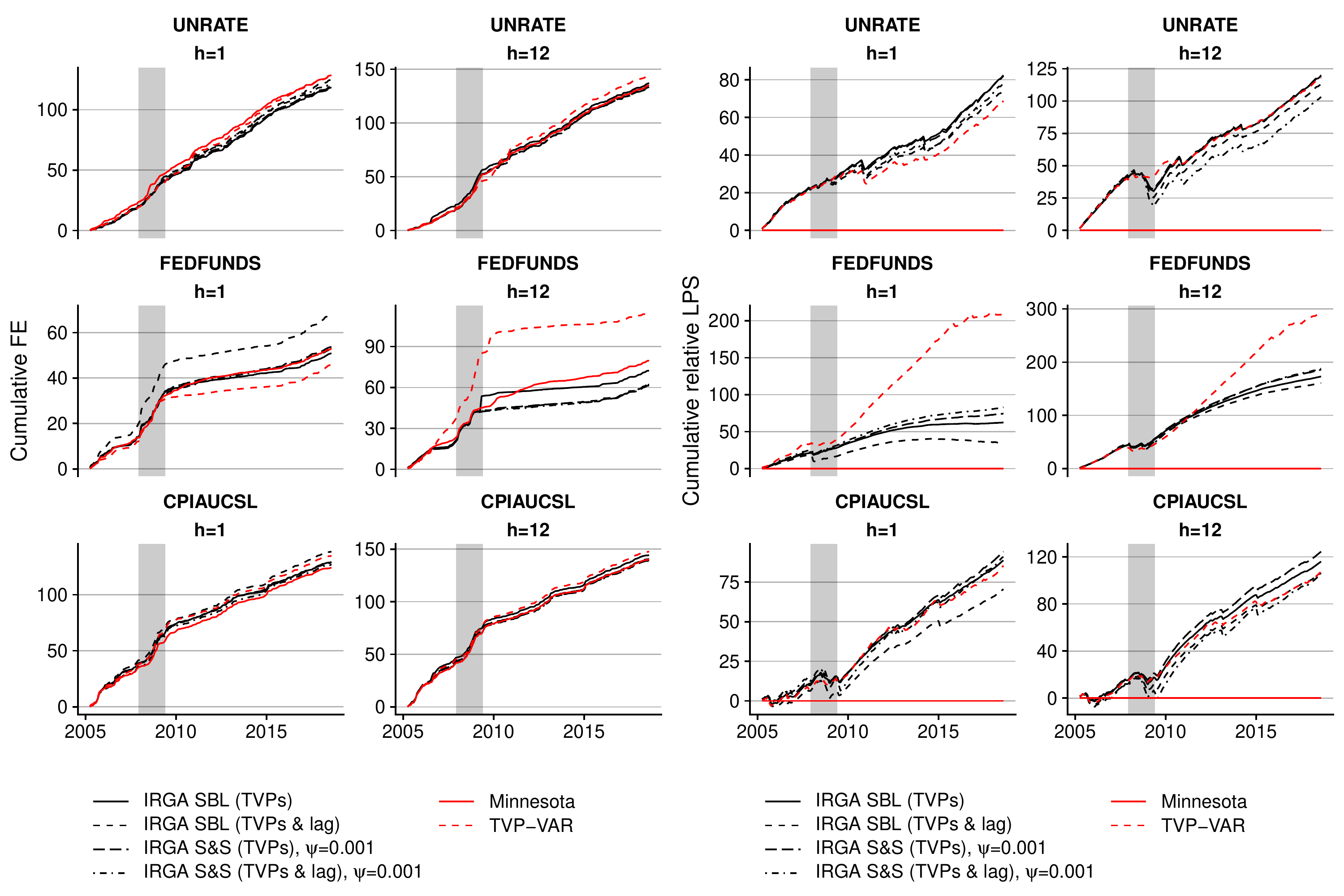}
\caption{Absolute cumulative forecast errors and log predictive scores relative to the benchmark over the holdout period 2005--2019.}\label{fig:forecasts}\vspace*{-0.3cm}
\caption*{\footnotesize\textit{Notes}: The grey shaded area indicates the Great Recession.}
\end{figure}

Figure \ref{fig:forecasts}, which plots our forecast metrics against time, tells of a story of constant improvements, since the Great Recession ended, in LPLs of our IRGA methods relative to the benchmark Bayesian VAR. Above, we noted the poor point forecast performance of the small TVP-VAR for the effective Federal funds rate for $h=12$. However, this model yielded good density forecasts for this case. Figure \ref{fig:forecasts}, along with the fact that the Federal funds rate was essentially zero for 2009-2016, suggests a possible explanation for this finding. The three-dimensional TVP-VAR is more parsimonious than any of the other models considered which all involved $78$ variables. Parsimonious models often lead to low predictive variances. In the long period when the Federal funds rate was constant, this property benefitted the small TVP-VAR. However, during the Great Recession, it performed much more poorly, as evidenced by the large forecast errors during this period.   

\begin{table*}[h!]
\caption{Forecast root mean squared error (RMSE) and average log predictive score (LPS) over the period 2005--2019.}\label{tab:fcst}\vspace*{-1.3em}
\begin{center}
\begin{threeparttable}
\footnotesize
\begin{tabular*}{\textwidth}{@{\extracolsep{\fill}} lrrrr}
\toprule
      & \multicolumn{2}{c}{\textbf{RMSE}} & \multicolumn{2}{c}{\textbf{average LPS}}\\
      \cmidrule(lr){2-3}\cmidrule(lr){4-5}
\textbf{Model} & $h=1$ & $h=12$ & $h=1$ & $h=12$\\ 
\midrule
  \textbf{Minnesota} &  &  &  &  \\ 
  UNRATE & 1.031 & 1.104 & -1.856 & -2.274 \\ 
  FEDFUNDS & 0.563 & 0.786 & -1.413 & -2.217 \\ 
  CPIAUCSL & 1.151 & 1.244 & -2.007 & -2.283 \\ 
  \midrule
  \textbf{IRGA SBL (TVPs)} &  &  &  &  \\ 
  UNRATE & 0.917 & 1.060 & 0.510 & 0.742 \\ 
  FEDFUNDS & 1.028 & 1.460 & 0.388 & 1.076 \\ 
  CPIAUCSL & 1.011 & 1.053 & 0.551 & 0.721 \\ 
  \textbf{IRGA SBL (TVPs \& lag)} &  &  &  &  \\ 
  UNRATE & 0.968 & 1.023 & 0.462 & 0.703 \\ 
  FEDFUNDS & 1.318 & 0.970 & 0.219 & 0.999 \\ 
  CPIAUCSL & 1.071 & 1.002 & 0.438 & 0.660 \\ 
  \textbf{IRGA S\&S (TVPs), $\psi=0.001$} &  &  &  &  \\ 
  UNRATE & 0.918 & 1.018 & 0.512 & 0.745 \\ 
  FEDFUNDS & 1.048 & 0.962 & 0.463 & 1.156 \\ 
  CPIAUCSL & 1.015 & 1.008 & 0.585 & 0.773 \\ 
  \textbf{IRGA S\&S (TVPs \& lag), $\psi=0.001$} &  &  &  &  \\ 
  UNRATE & 0.945 & 1.020 & 0.486 & 0.641 \\ 
  FEDFUNDS & 1.000 & 0.970 & 0.515 & 1.167 \\ 
  CPIAUCSL & 1.011 & 1.001 & 0.564 & 0.656 \\ 
  \midrule
  \textbf{TVP-VAR} &  &  &  &  \\ 
  UNRATE & 1.002 & 1.053 & 0.427 & 0.733 \\ 
  FEDFUNDS & 1.001 & 1.814 & 1.299 & 1.811 \\ 
  CPIAUCSL & 1.111 & 1.042 & 0.530 & 0.666 \\ 
\bottomrule
\end{tabular*}
\begin{tablenotes}[para,flushleft]
\scriptsize{\textit{Notes}: RMSEs and LPSs for the benchmark Bayesian VAR with hierarchical Minnesota prior, ratios for RMSEs and differences in LPSs for all others. RMSE entries less than unity indicate better performance than the benchmark for point forecasts, LPSs greater than zero indicate that the model has a better performance for density forecasts. SBL refers to the sparse Bayesian learning prior, S\&S is the spike-and-slab prior with $\psi$ performing best on average (additional results are in the Appendix). \textit{IRGA TVPs \& lag} is a specification where not only the TVPs are approximated, but also constant parameters for the second order cross-variable lags.}
\end{tablenotes}
\end{threeparttable}
\end{center}
\end{table*}

\subsection{Structural Analysis: Uncertainty Shocks}
The effect uncertainty shocks have on the economy is an important issue in modern macroeconomics. In this section, we update the seminal analysis of \cite{bloom2009} using our IRGA methods and various data sets taken from the large data set used in our forecasting exercise. The impulse responses to the uncertainty shock are identified using standard Cholesky identification with the uncertainty index (VXOCLSx) ordered first. We consider small, medium and large TVP-VARs. We use data sets involving four variables (the three core variables plus the uncertainty shock) as the small information set, and include additional series for medium (nine additional series) and large ($31$ additional variables). Table \ref{data_table} lists the variables included in each data set. For the structural analysis, we approximate the TVPs using VAMP relying on the SBL prior (results using the S\&S prior are very similar). The remaining parameters are estimated using MCMC methods.

\begin{figure}[ht]
\includegraphics[width=\textwidth]{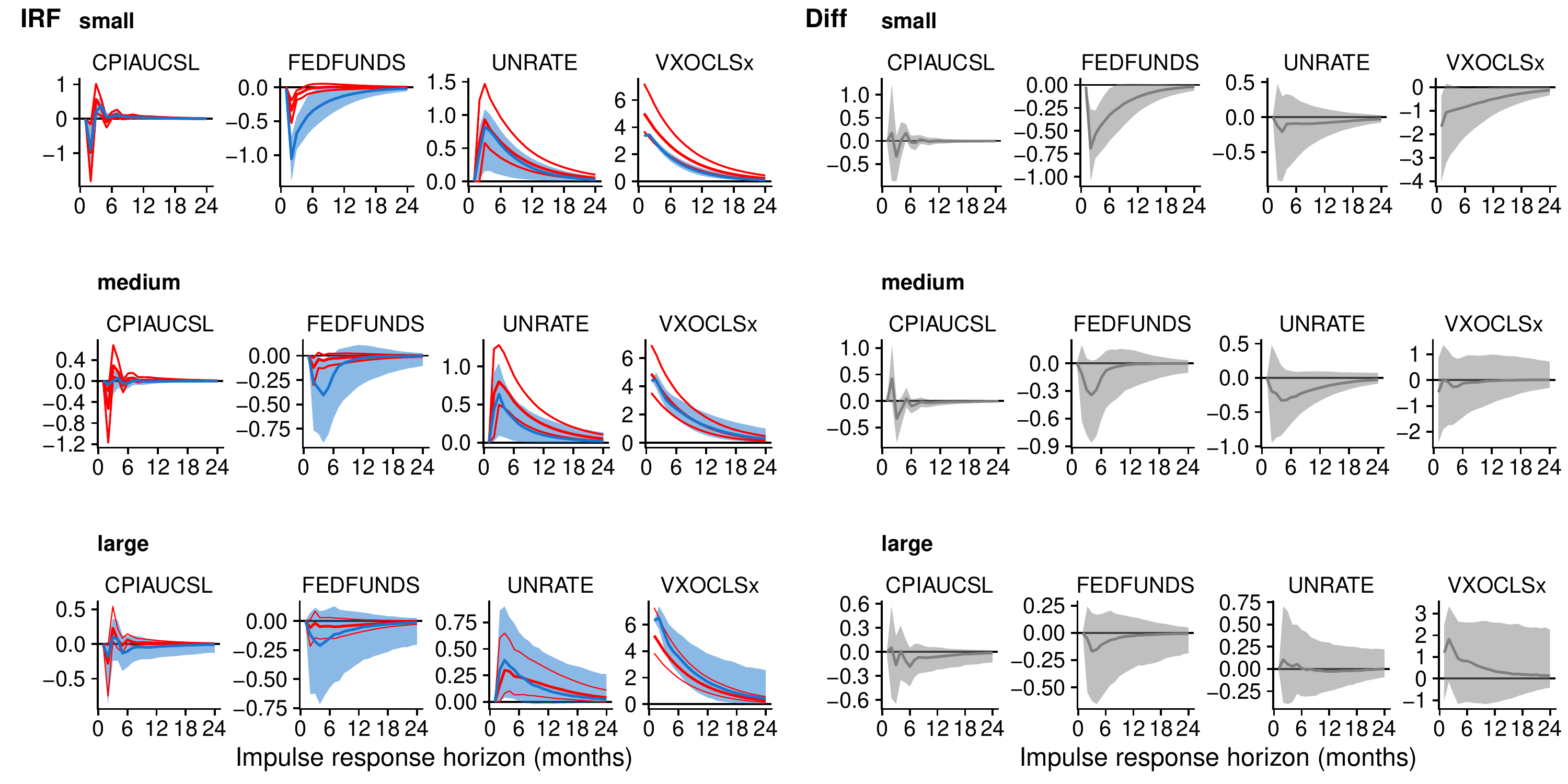}
\caption{Impulse response functions and differences between time-varying and constant parameter specification across model sizes.}\label{fig:irfs}\vspace*{-0.3cm}
\caption*{\footnotesize\textit{Notes}: See Tab. \ref{tab:data} in the Appendix for further details on the respective information sets. The blue area refers to the 16th and 84th credible sets of the IRFs associated with the IRGA approach (with the SBL prior) whereas the red lines are the IRFs associated with the constant parameter VAR. }
\end{figure}

With a TVP model, impulse responses vary over time. We present impulse responses using the constant coefficients. Hence, they can be interpreted as average or mean impulse responses. Figure \ref{fig:irfs} shows the impulse responses from a constant parameter VAR (in red, with $68$ percent posterior credible sets) versus our IRGA approach (blue, alongside $68$ percent posterior credible sets). The constant parameter VAR is the same as the one used as a benchmark in the forecasting section of the paper. The right panel plots posterior features of the difference in impulse responses between the constant parameter and TVP models. 

In general, results in the left panel of the figure are sensible. The uncertainty shock causes a substantial increase in unemployment and is associated with a drop in interest rates. Inflation falls on impact but then rises. The constant parameter and TVP models are both producing qualitatively similar findings. However, an examination of the right panel reveals an interesting pattern. When working with the large VAR, the constant parameter and TVP model are producing statistically similar impulse responses. That is, the posterior credible intervals for the difference in impulse responses between the two models always include zero. Thus, in the large model the inclusion of TVPs is having little or no impact on the impulse responses. However, for the small and (to a lesser extent) medium-sized VARs, the posterior credible intervals for the difference in impulse responses frequently do not include zero. This suggest that there is a trade-off between TVP-VAR dimension and the presence of TVPs. In small models, where the risk of omitted variables is high, the addition of TVPs clearly makes a difference. But in large models, where the risk is lower, the evidence for TVPs diminishes. 

Overall, we are finding that our IRGA approach is producing sensible results and that, by allowing for TVPs, we are obtaining some insights that would not be found in constant coefficient models. And it is worth stressing that IRGA methods are computationally effiicient, capable of handling very large TVP-VAR dimensions, where traditional MCMC-based TVP-VAR methods would not be computationally practicable. 

\section{Conclusions}
This paper has argued that, when working with TVP regressions with large data sets, care has to be taken to ensure that estimation methods are computationally feasible and that the appropriate type of time-variation is allowed for. That is, particularly when $K$ is large, it is likely that time-variation in regression coefficients only occurs for a few of the coefficients and only at some periods in time. With these considerations, we have developed computationally-efficient estimation methods which are scaleable and can accommodate a range of shrinkage priors that allow for great flexibility in the location and timing of parameter change. These are based on IRGAs. In artificial data, we document the computational efficiency and accuracy of these methods. In a large macroeconomic data set, we demonstrate good forecast performance in a practical amount of computation time. In an investigation of the role of uncertainty shocks, we show that TVP models yield insights that constant coefficient models do not. 

\small{
\addcontentsline{toc}{section}{References}
\bibliographystyle{./custom}
\bibliography{./tvp}}

\begin{thebibliography}{41}
\newcommand{\enquote}[1]{``#1''}
\providecommand{\natexlab}[1]{#1}

\bibitem[{Belmonte \emph{et~al.}(2014)Belmonte, Koop, and Korobilis}]{bkk}
\textsc{Belmonte M, Koop G, and Korobilis D} (2014), \enquote{Hierarchical
  shrinkage in time-varying coefficient models,} \emph{Journal of Forecasting}
  \textbf{33}(1), 80--94.

\bibitem[{Bhattacharya \emph{et~al.}(2015)Bhattacharya, Pati, Pillai, and
  Dunson}]{bppd2015}
\textsc{Bhattacharya A, Pati D, Pillai N, and Dunson D} (2015),
  \enquote{Dirichlet--Laplace priors for optimal shrinkage,} \emph{Journal of
  the American Statistical Association} \textbf{110}, 1479--1490.

\bibitem[{Bitto and Fr{\"u}hwirth-Schnatter(2019)}]{bitto_fs}
\textsc{Bitto A, and Fr{\"u}hwirth-Schnatter S} (2019), \enquote{Achieving
  shrinkage in a time-varying parameter model framework,} \emph{Journal of
  Econometrics} \textbf{210}(1), 75--97.

\bibitem[{Bloom(2009)}]{bloom2009}
\textsc{Bloom N} (2009), \enquote{The impact of uncertainty shocks,}
  \emph{Econometrica} \textbf{77}, 623--685.

\bibitem[{Carriero \emph{et~al.}(forthcoming)Carriero, Clark, and
  Marcellino}]{ccm2019}
\textsc{Carriero A, Clark T, and Marcellino M} (forthcoming), \enquote{Large
  Vector Autoregressions with stochastic volatility and flexible priors,}
  \emph{Journal of Econometrics} .

\bibitem[{Carter and Kohn(1994)}]{carter1994gibbs}
\textsc{Carter CK, and Kohn R} (1994), \enquote{On Gibbs sampling for state
  space models,} \emph{Biometrika} \textbf{81}(3), 541--553.

\bibitem[{Chan \emph{et~al.}(2012)Chan, Koop, Leon-Gonzalez, and
  Strachan}]{ckls2012}
\textsc{Chan J, Koop G, Leon-Gonzalez R, and Strachan R} (2012), \enquote{Time
  varying dimension models,} \emph{Journal of Business and Economic Statistics}
  \textbf{30}, 358--367.

\bibitem[{Clark(2011)}]{clark2011}
\textsc{Clark T} (2011), \enquote{Real-time density forecasts from BVARs with
  stochastic volatility,} \emph{Journal of Business and Economic Statistics}
  \textbf{29}, 327--341.

\bibitem[{Cogley and Sargent(2005)}]{cogley2005drifts}
\textsc{Cogley T, and Sargent TJ} (2005), \enquote{Drifts and volatilities:
  monetary policies and outcomes in the post WWII US,} \emph{Review of Economic
  Dynamics} \textbf{8}(2), 262 -- 302.

\bibitem[{D'Agostino \emph{et~al.}(2013)D'Agostino, Gambetti, and
  Giannone}]{giannone2013}
\textsc{D'Agostino A, Gambetti L, and Giannone D} (2013),
  \enquote{Macroeconomic forecasting and structural change,} \emph{Journal of
  Applied Econometrics} \textbf{28}(1), 82--101.

\bibitem[{Dangl and Halling(2012)}]{dh2012}
\textsc{Dangl T, and Halling M} (2012), \enquote{Predictive regressions with
  time-varying coefficients,} \emph{Journal of Financial Economics}
  \textbf{106}, 157--181.

\bibitem[{Fang \emph{et~al.}(2016)Fang, Zhang, and Li}]{fang2016two}
\textsc{Fang J, Zhang L, and Li H} (2016), \enquote{Two-dimensional
  pattern-coupled sparse Bayesian learning via generalized approximate message
  passing,} \emph{IEEE Transactions on Image Processing} \textbf{25}(6),
  2920--2930.

\bibitem[{Feldkircher \emph{et~al.}(2017)Feldkircher, Huber, and
  Kastner}]{feldkircher2017sophisticated}
\textsc{Feldkircher M, Huber F, and Kastner G} (2017), \enquote{Sophisticated
  and small versus simple and sizeable: When does it pay off to introduce
  drifting coefficients in Bayesian VARs?} \emph{arXiv} \textbf{1711.00564}.

\bibitem[{Giannone \emph{et~al.}(2015)Giannone, Lenza, and
  Primiceri}]{giannone2015prior}
\textsc{Giannone D, Lenza M, and Primiceri GE} (2015), \enquote{Prior selection
  for vector autoregressions,} \emph{Review of Economics and Statistics}
  \textbf{97}(2), 436--451.

\bibitem[{Griffin and Brown(2010)}]{griffin2010}
\textsc{Griffin J, and Brown P} (2010), \enquote{Inference with normal-gamma
  prior distributions in regression problems,} \emph{Bayesian Analysis}
  \textbf{5}(1), 171--188.

\bibitem[{Groen \emph{et~al.}(2013)Groen, Paap, and Ravazzollo}]{gpr2013}
\textsc{Groen J, Paap R, and Ravazzollo F} (2013), \enquote{Real time inflation
  forecasting in a changing world,} \emph{Journal of Business and Economic
  Statistics} \textbf{31}, 29--44.

\bibitem[{Hauzenberger \emph{et~al.}(2019)Hauzenberger, Huber, Koop, and
  Onorante}]{hauzenberger2019fast}
\textsc{Hauzenberger N, Huber F, Koop G, and Onorante L} (2019), \enquote{Fast
  and Flexible Bayesian Inference in Time-varying Parameter Regression Models,}
  \emph{arXiv} \textbf{1910.10779}.

\bibitem[{Huber \emph{et~al.}(forthcoming)Huber, Koop, and Onorante}]{hko2019}
\textsc{Huber F, Koop G, and Onorante L} (forthcoming), \enquote{Inducing
  sparsity and shrinkage in time-varying parameter models,} \emph{Journal of
  Business and Economic Statistics} .

\bibitem[{Kalli and Griffin(2014)}]{kg2014}
\textsc{Kalli M, and Griffin J} (2014), \enquote{Time-varying sparsity in
  dynamic regression models,} \emph{Journal of Econometrics} \textbf{178},
  779--793.

\bibitem[{Kastner and Fr{\"u}hwirth-Schnatter(2014)}]{kastner2014ancillarity}
\textsc{Kastner G, and Fr{\"u}hwirth-Schnatter S} (2014),
  \enquote{Ancillarity-sufficiency interweaving strategy (ASIS) for boosting
  MCMC estimation of stochastic volatility models,} \emph{Computational
  Statistics \& Data Analysis} \textbf{76}, 408--423.

\bibitem[{Kastner and Huber(2017)}]{kastnerhuber2017}
\textsc{Kastner G, and Huber F} (2017), \enquote{Sparse Bayesian vector
  autoregressions in huge dimensions,} \emph{manuscript} .

\bibitem[{Kim \emph{et~al.}(1998)Kim, Shephard, and Chib}]{kim1998stochastic}
\textsc{Kim S, Shephard N, and Chib S} (1998), \enquote{Stochastic volatility:
  likelihood inference and comparison with ARCH models,} \emph{The Review of
  Economic Studies} \textbf{65}(3), 361--393.

\bibitem[{Koop and Korobilis(2012)}]{kk2012}
\textsc{Koop G, and Korobilis D} (2012), \enquote{Forecasting inflation using
  dynamic model averaging,} \emph{International Economic Review}
  \textbf{53}(3), 867--886.

\bibitem[{Koop and Korobilis(2013)}]{koop2013large}
---{}---{}--- (2013), \enquote{Large time-varying parameter VARs,}
  \emph{Journal of Econometrics} \textbf{177}(2), 185--198.

\bibitem[{Koop and Korobilis(2018)}]{koopkorobilisvb}
---{}---{}--- (2018), \enquote{Variational Bayes inference in high dimensional
  time-varying parameter models,} \emph{manuscript} .

\bibitem[{Koop \emph{et~al.}(2019)Koop, Korobilis, and Pettenuzzo}]{kkp2019}
\textsc{Koop G, Korobilis D, and Pettenuzzo D} (2019), \enquote{Bayesian
  compressed Vector Autoregressions,} \emph{Journal of Econometrics}
  \textbf{210}(1), 135--154.

\bibitem[{Korobilis(2019)}]{korobilis2019high}
\textsc{Korobilis D} (2019), \enquote{High-dimensional macroeconomic
  forecasting using message passing algorithms,} \emph{Journal of Business \&
  Economic Statistics} \textbf{forthcoming}, 1--30.

\bibitem[{Kowal \emph{et~al.}(2017)Kowal, Matteson, and Ruppert}]{kmr2017}
\textsc{Kowal D, Matteson D, and Ruppert D} (2017), \enquote{Dynamic shrinkage
  processes,} \emph{arXiv:1707.00763} .

\bibitem[{Litterman(1986)}]{litterman1986forecasting}
\textsc{Litterman RB} (1986), \enquote{Forecasting with Bayesian vector
  autoregressions—five years of experience,} \emph{Journal of Business \&
  Economic Statistics} \textbf{4}(1), 25--38.

\bibitem[{McCracken and Ng(2016)}]{mccracken2016fredmd}
\textsc{McCracken MW, and Ng S} (2016), \enquote{FRED-MD: A monthly database
  for macroeconomic research,} \emph{Journal of Business \& Economic
  Statistics} \textbf{34}(4), 574--589.

\bibitem[{Mitchell and Beauchamp(1988)}]{mitchell1988bayesian}
\textsc{Mitchell TJ, and Beauchamp JJ} (1988), \enquote{Bayesian variable
  selection in linear regression,} \emph{Journal of the American Statistical
  Association} \textbf{83}(404), 1023--1032.

\bibitem[{Nakajima and West(2013)}]{nw2013}
\textsc{Nakajima J, and West M} (2013), \enquote{Bayesian analysis of latent
  threshold dynamic models,} \emph{Journal of Business and Economic Statistics}
  \textbf{31}, 151–164.

\bibitem[{Paul(2019)}]{paul2019timevarying}
\textsc{Paul P} (2019), \enquote{The time-varying effect of monetary policy on
  asset prices,} \emph{Review of Economics and Statistics}
  \textbf{forthcoming}, 1--44.

\bibitem[{Primiceri(2005)}]{primiceri2005}
\textsc{Primiceri G} (2005), \enquote{Time varying structural autoregressions
  and monetary policy,} \emph{Oxford University Press} \textbf{72}(3),
  821--852.

\bibitem[{Rangan \emph{et~al.}(2019)Rangan, Schniter, and
  Fletcher}]{rangan2019vector}
\textsc{Rangan S, Schniter P, and Fletcher AK} (2019), \enquote{Vector
  approximate message passing,} \emph{IEEE Transactions on Information Theory}
  \textbf{65}(10), 6664--6684.

\bibitem[{Ro\v{c}kov\'{a} and McAlinn(2018)}]{rm2018}
\textsc{Ro\v{c}kov\'{a} V, and McAlinn K} (2018), \enquote{Dynamic variable
  selection with spike-and-slab process priors,} \emph{Technical report, Booth
  School of Business, University of Chicago} .

\bibitem[{Stock and Watson(2007)}]{sw2007}
\textsc{Stock J, and Watson M} (2007), \enquote{Why has U.S. inflation become
  harder to forecast?} \emph{Journal of Money, Credit and Banking} \textbf{39},
  3--33.

\bibitem[{Tipping(2001)}]{tipping2001sparse}
\textsc{Tipping ME} (2001), \enquote{Sparse Bayesian learning and the relevance
  vector machine,} \emph{Journal of machine learning research} \textbf{1}(Jun),
  211--244.

\bibitem[{Uribe and Lopes(2017)}]{ul2017}
\textsc{Uribe P, and Lopes H} (2017), \enquote{Dynamic sparsity on dynamic
  regression models,} \emph{manuscript} .

\bibitem[{{van den Boom} \emph{et~al.}(2019){van den Boom}, Reeves, and
  Dunson}]{boom2019approximating}
\textsc{{van den Boom} W, Reeves G, and Dunson DB} (2019),
  \enquote{Approximating posteriors with high-dimensional nuisance parameters
  via integrated rotated Gaussian approximation,} \emph{arXiv}
  \textbf{1909.06753}.

\bibitem[{Zou \emph{et~al.}(2016)Zou, Li, Fang, and
  Li}]{zou2016computationally}
\textsc{Zou X, Li F, Fang J, and Li H} (2016), \enquote{Computationally
  efficient sparse Bayesian learning via generalized approximate message
  passing,} in \enquote{2016 IEEE International Conference on Ubiquitous
  Wireless Broadband (ICUWB),} 1--4, IEEE.

\end{thebibliography}

\newpage
\begin{appendices}
\setcounter{equation}{0}
\renewcommand\theequation{A.\arabic{equation}}
\section{A TVP-VAR model}\label{sec: appVAR}
In this appendix, we briefly show how our methods can be applied to the VAR case.
Let $\bm Y$ denote a $T \times N$-dimensional matrix of endogenous variables with $t^{th}$ row given by $\bm y'_t = (y_{1t}, \dots, y_{Nt})$. Using this notation, the TVP-VAR is given by:
\begin{equation}
\bm y_t = (\bm I_t \otimes \bm x'_t) (\bm \beta + \bm \gamma_t) + \bm \varepsilon_t. \label{eq: VAR}
\end{equation} 
Here,  $\bm x_t = (\bm y'_{t-1}, \dots, \bm y'_{t-P})'$ now includes the $P$ lags of $\bm y_t$ and $\bm \varepsilon_t$ is a $N$-dimensional Gaussian shock vector with time-varying variance-covariance matrix given by $\bm \Sigma_t$ of dimension $N \times N$. 

Equation \ref{eq: VAR} can be rewritten on an equation-by-equation basis by augmenting $\bm x_t$ in the $j^{th}$ equation (for $j \ge 2$) of the system with the  contemporaneous values  of the first $j-1$ elements of $\bm y_t$:
\begin{equation*}
 y_{jt} =  \tilde{\bm \beta}'_j\tilde{\bm x}_{jt}  + \tilde{\bm \gamma}'_{jt} \tilde{\bm x}_{jt}  + \bm \varepsilon_{jt}, \label{eq: regmodel}
\end{equation*}
 whereby $\tilde{\bm x}_{jt}= (\bm x'_t,  y_{1t}, \dots, y_{j-1, t})'$, $\tilde{\bm \beta}_j$ is a $K_j (= Np + (j-1))$-dimensional vector of equation-specific regression coefficients and $\tilde{\bm \gamma}_{jt}$ denotes an $K_j \times 1$ vector of TVPs specific to equation $j$. The last $j-1$ elements in $ \tilde{\bm \beta}_j$ and $\tilde{\bm \gamma}_{jt}$ serve to approximate the covariance parameters and their time variation in $\bm \Sigma_t$.  In the case that $j=1$, $\tilde{\bm x}_{jt}= \bm x_t$.  
 
 Note that the equation-by-equation form of the TVP-VAR has great advantages in terms of computation since the computational burden of estimating a full system can be reduced efficiently by exploiting the fact that (\ref{eq: regmodel}) constitutes a set of $N$ independent regression models that can be estimated on a cluster.

\section{Data and results}\label{app:dataresults}
In this appendix, we repeat the forecasting exercise using the S\&S prior for a range of values for the key prior hyperparameter $\psi$. It can be see that results are very robust to prior choice.
\begin{table*}[h!]
\caption{Forecast RMSE and average LPS for the S\&S prior with different values of $\psi$.}\label{tab:fcst_psi}\vspace*{-1.5em}
\begin{center}
\begin{threeparttable}
\footnotesize
\begin{tabular*}{\textwidth}{@{\extracolsep{\fill}} lrrrr}
\toprule
      & \multicolumn{2}{c}{\textbf{RMSE}} & \multicolumn{2}{c}{\textbf{average LPS}}\\
      \cmidrule(lr){2-3}\cmidrule(lr){4-5}
\textbf{Model} & $h=1$ & $h=12$ & $h=1$ & $h=12$\\ 
\midrule
  \textbf{Minnesota} &  &  &  &  \\ 
  UNRATE & 1.031 & 1.104 & -1.86 & -2.27 \\ 
  FEDFUNDS & 0.563 & 0.786 & -1.41 & -2.22 \\ 
  CPIAUCSL & 1.151 & 1.244 & -2.01 & -2.28 \\ 
  \midrule
  \textbf{IRGA S\&S (TVPs), $\psi=1e-06$} &  &  &  &  \\ 
  UNRATE & 0.915 & 1.012 & 0.513 & 0.729 \\ 
  FEDFUNDS & 1.022 & 0.946 & 0.417 & 1.099 \\ 
  CPIAUCSL & 1.009 & 1.003 & 0.557 & 0.715 \\ 
  \textbf{IRGA S\&S (TVPs), $\psi=1e-04$} &  &  &  &  \\ 
  UNRATE & 0.917 & 1.016 & 0.512 & 0.729 \\ 
  FEDFUNDS & 1.018 & 0.969 & 0.425 & 1.104 \\ 
  CPIAUCSL & 1.007 & 1.005 & 0.562 & 0.722 \\ 
  \textbf{IRGA S\&S (TVPs), $\psi=0.001$} &  &  &  &  \\ 
  UNRATE & 0.918 & 1.018 & 0.512 & 0.745 \\ 
  FEDFUNDS & 1.048 & 0.962 & 0.463 & 1.156 \\ 
  CPIAUCSL & 1.015 & 1.008 & 0.585 & 0.773 \\ 
  \textbf{IRGA S\&S (TVPs), $\psi=0.01$} &  &  &  &  \\ 
  UNRATE & 0.934 & 1.276 & 0.439 & 0.638 \\ 
  FEDFUNDS & 1.135 & 1.337 & 0.068 & 0.686 \\ 
  CPIAUCSL & 1.059 & 1.234 & 0.331 & 0.433 \\ 
  \textbf{IRGA S\&S (TVPs), $\psi=1/K$} &  &  &  &  \\ 
  UNRATE & 0.923 & 1.025 & 0.509 & 0.745 \\ 
  FEDFUNDS & 1.122 & 0.978 & 0.248 & 0.911 \\ 
  CPIAUCSL & 1.052 & 1.022 & 0.435 & 0.574 \\ 
  \textbf{IRGA S\&S (TVPs \& lag), $\psi=1e-06$} &  &  &  &  \\ 
  UNRATE & 0.958 & 1.021 & 0.472 & 0.680 \\ 
  FEDFUNDS & 1.267 & 0.967 & 0.278 & 1.032 \\ 
  CPIAUCSL & 1.059 & 1.002 & 0.458 & 0.646 \\ 
  \textbf{IRGA S\&S (TVPs \& lag), $\psi=1e-04$} &  &  &  &  \\ 
  UNRATE & 0.957 & 1.019 & 0.473 & 0.668 \\ 
  FEDFUNDS & 1.215 & 0.964 & 0.317 & 1.057 \\ 
  CPIAUCSL & 1.050 & 1.002 & 0.472 & 0.643 \\ 
  \textbf{IRGA S\&S (TVPs \& lag), $\psi=0.001$} &  &  &  &  \\ 
  UNRATE & 0.945 & 1.020 & 0.486 & 0.641 \\ 
  FEDFUNDS & 1.000 & 0.970 & 0.515 & 1.167 \\ 
  CPIAUCSL & 1.011 & 1.001 & 0.564 & 0.656 \\ 
  \textbf{IRGA S\&S (TVPs \& lag), $\psi=0.01$} &  &  &  &  \\ 
  UNRATE & 0.964 & 1.014 & 0.360 & 0.673 \\ 
  FEDFUNDS & 0.941 & 0.977 & 0.015 & 0.720 \\ 
  CPIAUCSL & 1.063 & 1.013 & 0.315 & 0.527 \\ 
  \textbf{IRGA S\&S (TVPs \& lag), $\psi=1/K$} &  &  &  &  \\ 
  UNRATE & 0.963 & 1.021 & 0.441 & 0.744 \\ 
  FEDFUNDS & 0.947 & 0.971 & 0.213 & 0.916 \\ 
  CPIAUCSL & 1.057 & 1.002 & 0.434 & 0.644 \\ 
\bottomrule
\end{tabular*}
\begin{tablenotes}[para,flushleft]
\scriptsize{\textit{Notes}: RMSEs and LPSs for the benchmark Bayesian VAR with hierarchical Minnesota prior, ratios for RMSEs and differences in LPSs for all others. RMSE entries less than unity indicate better performance than the benchmark for point forecasts, LPSs greater than zero indicate that the model has a better performance for density forecasts. S\&S is the spike-and-slab prior. \textit{IRGA TVPs \& lag} is a specification where not only the TVPs are approximated, but also constant parameters for the second order cross-variable lags.}
\end{tablenotes}
\end{threeparttable}
\end{center}
\end{table*}

\begin{landscape}
\begin{table*}[ht]
\caption{Monthly data obtained from FRED-MD for 1985:01 to 2019:08.}\label{tab:data}\vspace*{-1.5em}
\begin{center}
\begin{threeparttable}
\scriptsize
\begin{tabular*}{25cm}{@{\extracolsep{\fill}} lclcl|lclc}
  \toprule
Mnemonic & $I(0)$ & Description & Size & & Mnemonic & $I(0)$ & Description & Size \\ 
  \midrule
RPI & 5 & Real personal income &  &  & ANDENOx & 5 & New Orders for Nondefense Capital goods &  \\ 
  W875RX1 & 5 & Real personal income ex transfer receipts &  &  & AMDMUOx & 5 & Unfilled Orders for Durable goods &  \\ 
  INDPRO & 5 & IP Index & M &  & UMCSENTx & 2 & Consumer Sentiment Index & L \\ 
  IPFPNSS & 5 & IP: Final Products &  &  & BUSLOANS & 6 & Commercial and Industrial Loans & L \\ 
  IPFINAL & 5 & IP: Final Products (Market Group) &  &  & REALLN & 6 & Real Estate Loans at All Commerical Banks & L \\ 
  IPCONGD & 5 & IP: Consumer Goods &  &  & INVEST & 6 & Securities in Bank Credit at All Commercial Banks & M \\ 
  IPDCONGD & 5 & IP: Durable Consumer Goods &  &  & FEDFUNDS & 2 & Effective Federal Funds Rate & S \\ 
  IPNCONGD & 5 & IP: Nondurable Consumer Goods &  &  & CP3Mx & 2 & 3-Month AA Financial Commercial Paper Rate & L \\ 
  IPBUSEQ & 5 & IP: Business Equipment &  &  & TB3MS & 2 & 3-Month Treasury Bill & L \\ 
  IPMAT & 5 & IP: Materials &  &  & TB6MS & 2 & 6-Month Treasury Bill & L \\ 
  IPDMAT & 5 & IP: Durable Materials &  &  & GS1 & 2 & 1-Year Treasury Rate & L \\ 
  IPNMAT & 5 & IP: Nondurable Materials &  &  & GS5 & 2 & 5-Year Treasury Rate & L \\ 
  IPMANSICS & 5 & IP: Manufacturing (SIC) &  &  & GS10 & 2 & 10-Year Treasury Rate & M \\ 
  IPFUELS & 5 & IP: Fuels &  &  & AAA & 2 & Moody's Seasoned Aaa Corporate Bond Yield & M \\ 
  CUMFNS & 2 & Capacity Utilization: Manufacturing &  &  & BAA & 2 & Moody's Seasoned Baa Corporate Bond Yield & M \\ 
  UNRATE & 2 & Civilian Unemployment Rate & S &  & COMPAPFFx & 1 & 3-Month Commercial Paper Minus FEDFUNDS &  \\ 
  UEMPMEAN & 2 & Average Duration of Unemployment (Weeks) & L &  & TB3SMFFM & 1 & 3-Month Treasury C  Minus FEDFUNDS &  \\ 
  UEMPLT5 & 5 & Civilians Unemployed: Less Than 5 Weeks &  &  & TB6SMFFM & 1 & 6-Month Treasury C  Minus FEDFUNDS &  \\ 
  UEMP5TO14 & 5 & Civilians Unemployed for 5-14 Weeks &  &  & T1YFFM & 1 & 1-Year Treasury C  Minus FEDFUNDS &  \\ 
  UEMP15OV & 5 & Civilians Unemployed: 15 Weeks and Over &  &  & T5YFFM & 1 & 5-Year Treasury C  Minus FEDFUNDS &  \\ 
  UEMP15T26 & 5 & Civilians Unemployed for 15-26 Weeks &  &  & T10YFFM & 1 & 10-Year Treasury C  Minus FEDFUNDS & M \\ 
  UEMP27OV & 5 & Civilians Unemployed for 27 Weeks and Over &  &  & AAAFFM & 1 & Moody's Aaa Corporate Bond  Minus FEDFUNDS &  \\ 
  CLAIMSx & 5 & Initial Claims & L &  & BAAFFM & 1 & Moody's Baa Corporate Bond  Minus FEDFUNDS &  \\ 
  PAYEMS & 5 & All Employees: Total nonfarm & L &  & TWEXMMTH & 5 & Trade Weighted Trade Weighted U.S. Dollar Index: Major Currencies &  \\ 
  USGOOD & 5 & All Employees: Goods-Producing Industries &  &  & EXSZUSx & 5 & Switzerland--U.S. Foreign Exchange Rate & L \\ 
  USCONS & 5 & All Employees: Construction &  &  & EXJPUSx & 5 & Japan--U.S. Foreign Exchange Rate & L \\ 
  MANEMP & 5 & All Employees: Manufacturing &  &  & EXUSUKx & 5 & U.S.--UK Foreign Exchange Rate & L \\ 
  SRVPRD & 5 & All Employees: Service-Providing Industries &  &  & EXCAUSx & 5 & Canada--U.S. Foreign Exchange Rate & L \\ 
  USTPU & 5 & All Employees: Trade, Transporation and Utilities &  &  & OILPRICEx & 6 & Crude Oil, , spliced WTI and Cushing & M \\ 
  USWTRADE & 5 & All Employees: Wholesale Trade &  &  & PPICMM & 6 & PPI: Metals and metal products &  \\ 
  USTRADE & 5 & All Employees: Retail Trade &  &  & CPIAUCSL & 6 & CPI: All Items & S \\ 
  USFIRE & 5 & All Employees: Financial Activities &  &  & CPIAPPSL & 6 & CPI: Apparel &  \\ 
  USGOVT & 5 & All Employees: Government &  &  & CPITRNSL & 6 & CPI: Transportation &  \\ 
  AWOTMAN & 2 & Avg Weekly Overtime Hourse: Manufacturing & M &  & CPIMEDSL & 6 & CPI: Medical Care &  \\ 
  AWHMAN & 1 & Avg Weekly Hours: Manufacturing & L &  & CPIULFSL & 6 & CPI: All Items Less Food &  \\ 
  HOUST & 4 & Housing Starts: Total New Privately Owned & M &  & PCEPI & 6 & Personal Cons. Expend.: Chain Index & M \\ 
  PERMIT & 4 & New Private Housing Permits (SAAR) & M &  & S\&P 500 & 5 & S\&P's Common Stock Price Index: Composite & M \\ 
  RETAILx & 5 & Retail and Food Services Sales & L &  & S\&P: indust & 5 & S\&P's Common Stock Price Index: Industrials & L \\ 
  AMDMNOx & 5 & New Orders for Durable goods &  &  & VXOCLSx & 1 & VXO & S \\ 
  \bottomrule
\end{tabular*}
\begin{tablenotes}[para,flushleft]
\scriptsize{\textit{Notes}: The dataset discussed in \citet{mccracken2016fredmd} is available for download at \href{https://fred.stlouisfed.org}{fred.stlouisfed.org}. The column $I(0)$ indicates the applied transformation to a series $x_t$ for obtaining stationary series: (1) no transformation, (2) $\Delta x_t$, (5) $\Delta \log(x_t)$, (6) $\Delta^2 \log(x_t)$ with $\Delta^i$ indicating $i$th differences. S (small), M (medium) and L (large) indicate inclusion in differently sized information sets for the structural analysis, with the letter referring to additional variables per class.}
\end{tablenotes}
\end{threeparttable}
\end{center}
\label{data_table}
\end{table*}
\end{landscape}
\end{appendices}
\end{document}